\documentclass[11pt,twoside]{article}

\usepackage{baaa-eng}
\usepackage{graphicx}
\usepackage{subfigure}
\usepackage{psfrag}
\usepackage{amssymb}

\usepackage[latin1]{inputenc}

\usepackage[T1]{fontenc} 

\usepackage{latexsym}
\usepackage{verbatim}
\usepackage[colorlinks=true,dvips]{hyperref}

\begin{document}
\vskip 1.0cm

\pagestyle{myheadings}
\vspace*{0.5cm}
\parindent 0pt{POSTER PAPER } 
\vskip 0.3cm

\title{Gemini GMOS IFU Spectroscopy of IRAS 04505-2958:
A New Exploding BAL + IR + Fe II QSO}

\author{S. Lipari$^1$, M. Bergmann$^2$, S. Sanchez$^3$, R. Terlevich$^4$,
Y.Taniguchi$^5$, E.Mediavilla$^6$, B.Punsly, B.Garcia$^6$,
W.Zheng$^7$, D.Merlo$^1$}

\affil{$^1$Observatorio Astronomico de Cordoba and CONICET;
$^2$Gemini Observatory, Chile;
$^3$Calar Alto Observatory, Spain;
$^4$Univ. of Cambridge, UK;
$^5$Tohoku Univ., Japan
$^6$Inst. de Astrof\'{\i}sica de Canarias, Spain;
$^7$Johns Hopkins Univ., USA;}

\begin{abstract}
New results of a Gemini GMOS Programme of study of  BAL+ IR+ Fe II QSOs
are presented.
We have performed a study of the kinematics, morphological, and
physical conditions,  in the BAL-QSO: IRAS 04505-2958.
From this study, selected results are presented,
mainly for the three more internal expanding shells.
In particular,
the GMOS data suggest that the outflow processes --in this QSO--
generated multiple  expanding hypergiant shells
(from $\sim$10, to $\sim$100 kpc), in several extreme explosive events.
These new GMOS data are in good agreement with our evolutionary, 
explosive and composite model: where part of the ISM of the host galaxy
is ejected in the form of multiple giant shells, mainly by HyN explosions.
This process could generate satellite/companion galaxies, and even
could expel a high fraction --or all-- the host galaxy. In addition,
this Model for AGN could give  important clues
about the origin --in AGNs-- of very energetic cosmic rays, observed by
the P. Auger Observatory

\end{abstract}

\begin{resumen}
Nuevos resultados de un programa Gemini GMOS de estudio de BAL + IR +
Fe II QSOs son presentados.
Nosotros estudiamos la morfologia, cinematica y condiciones fisicas de
IRAS 04505-2958.
Resultados selectos de este estudio se presentan en este trabajo,
sobre 3 de sus burbujas en expansion.
Estos datos muestran un buen acuerdo con nuesto Modelo evolutivo,
explosivo y compuesto de AGNs/QSOs. Este modelo explosivo de
AGN podria darnos claves sobre el origen --en AGN-- de los rayos cosmicos
detectados por el Obsvatorio P. Auger.

\end{resumen}


\section{Introduction, The Programme, and Observations}

There is increasing evidence that galactic outflow (OF), broad absorption
lines (BALs) and explosive events (ExE) play a main
role, specially when the galaxies and QSOs formed (see Lipari et al. 2007).


{\bf 1.1. Explosive BAL + IR + Fe II QSOs.}
The presence of {\it extreme explosions and OF}
--associated mostly to the end of extreme massive stars-- is  an important
component for different theoretical models of galaxy/QSO 
evolution (see Lipari et al. 2007).
From the observational point of view,
the presence of multiple concentric expanding supergiant bubbles/shells
in young composite BAL + IR + Fe II QSOs,
with centre in the nucleus and with highly symmetric circular shape could be
associated mainly with giant symmetric explosive events (L\'{i}pari et al
2003).
These explosive events could be explained in a composite
scenario: where  the interaction between the starburst and
the AGN could generate giant explosive/HyperNova (HyN) events
(see Collin \& Zahn 1999).


{\bf 1.2. Evolutionary, Explosive and Composite Model for QSOs/AGNs.}
An evolutionary, explosive and composite scenario was proposed for BAL + IR
+ Fe\,{\sc ii} QSOs (L\'{i}pari 1994, Lipari et al. 2005, Lipari \&
Terlevich 2006). Where mergers fuel extreme star formation
processes and  AGNs, resulting in strong dust and IR emission, large
number of SN and HyN events, with expanding super giant bubbles/shells.

{\bf 1.3. IRAS 04505-2958.}
This IR source was associated with a luminous quasar (M$_V =$ --25.8)
at z $=$ 0.286. The first
optical images and spectroscopy showed a bright nucleus, a close
foreground G star (at 2$''$ to the NW, from the nucleus) plus a
possible tidal tail to the SE (at $\sim$2$''$). 
HST WFPC2  images show that the possible SE
``tail'' is a very complex structure.
Lipari et al. (2003) suggested that the SE tail/ring like
structure is probably a large scale (30 kpc) external expanding shell,
at r $\sim$11 kpc.
The BAL system in this IR QSO was discovered by Lipari et al.(2005)
based on the evolutionary IR colour-colour diagram (Fig.\ 15 in their
paper). Notably, the BAL shows the same blueshift as the main OF.

{\bf 1.5. Gemini GMOS-IFU and HST Observations.}
This study is based on Gemini integral field spectroscopy,
combined with Hubble Space Telescope images.
3D  deep optical spectroscopy of  the nuclei and the
3 more internal arcs of IRAS~04505$-$2958
 were obtained during 4 nights in  2005 and 2007, at Gemini South.  
The telescope was used with the  Multi-Object Spectrograph (GMOS).

\section{The Hypergiant Shells System of IRAS 04505-2958:
Gemini + HST Evidence of Multiple Hyper-explosive Events.}

{\bf 2.1. Previous Works}.

{\bf Shells S1 and S2:}
From a study of host galaxies in QSOs, Magain et al. (2005) detected
a blob close to IRAS 04505-2958, about 0.3$''$
to the north-west; without other clear evidence of the host galaxy.
In the present GMOS study we found that this blob is composite by 2
shells (S1 and S2) of r  $\sim$0.2, 0.4$''$ (1.1, 2.2 kpc).

{\bf Shell S3:}
From a study of HST images + CASLEO spectra of 
IRAS~04505$-$2958, we proposed
that this extended/hyper shell was generated by explosive events with
{\bf composite hyperwinds} (Lipari et al. 2003, 2005).

{\bf Shell  S4:}
A study of a very clear external supergiant shell at r $\sim$ 15$''$
($\sim$80 kpc) is actually in progress.
The study of the very extended shells S4 and S3 is important for
the analysis of the origin of the extended Ly$\alpha$ blobs at
high redshift (Lipari \& Terlevich 2006).

{\bf 2.2. Gemini Results}.
Fig. 1a presents high resolution HST WFPC2   
broad-band image/contours obtained in the optical wavelengths through
the filter WF2-F702W. This HST image 
shows: the QSO, the main concentric shell S3,  and the field G star.
In addition, Figs. 1a shows 
the observed GMOS field ($\sim$20 kpc $\times$ 28 kpc).
The GMOS frame was centred close to the middle position between the QSO
and S3, at the PA $\sim$ 311$^o$.
Fig. 1b shows the [S {\sc ii}]/H$\alpha$ GMOS map.

\begin{figure}[!ht]
  \centering
  \includegraphics[width=.40\textwidth]{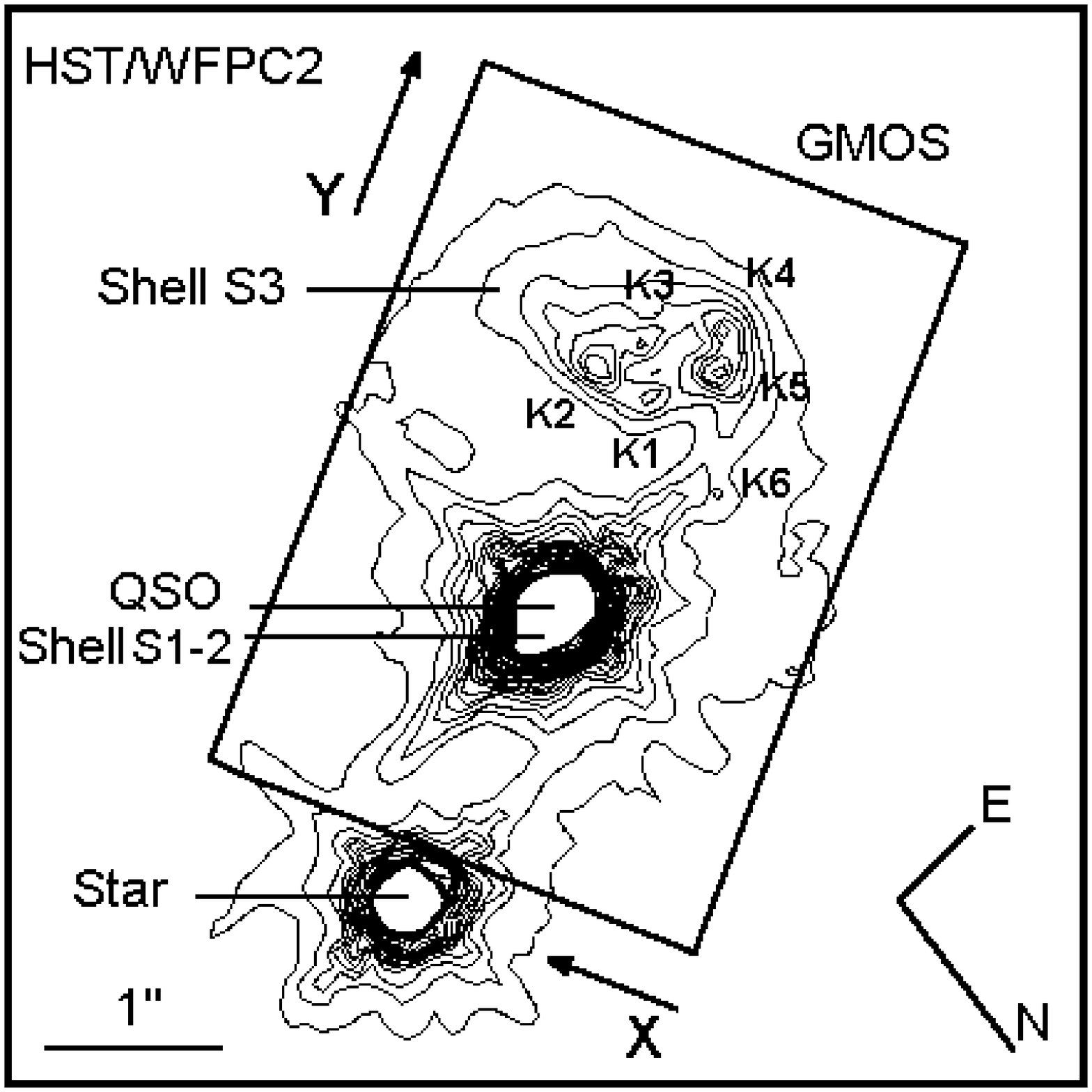}~\hfill
  \includegraphics[width=.40\textwidth]{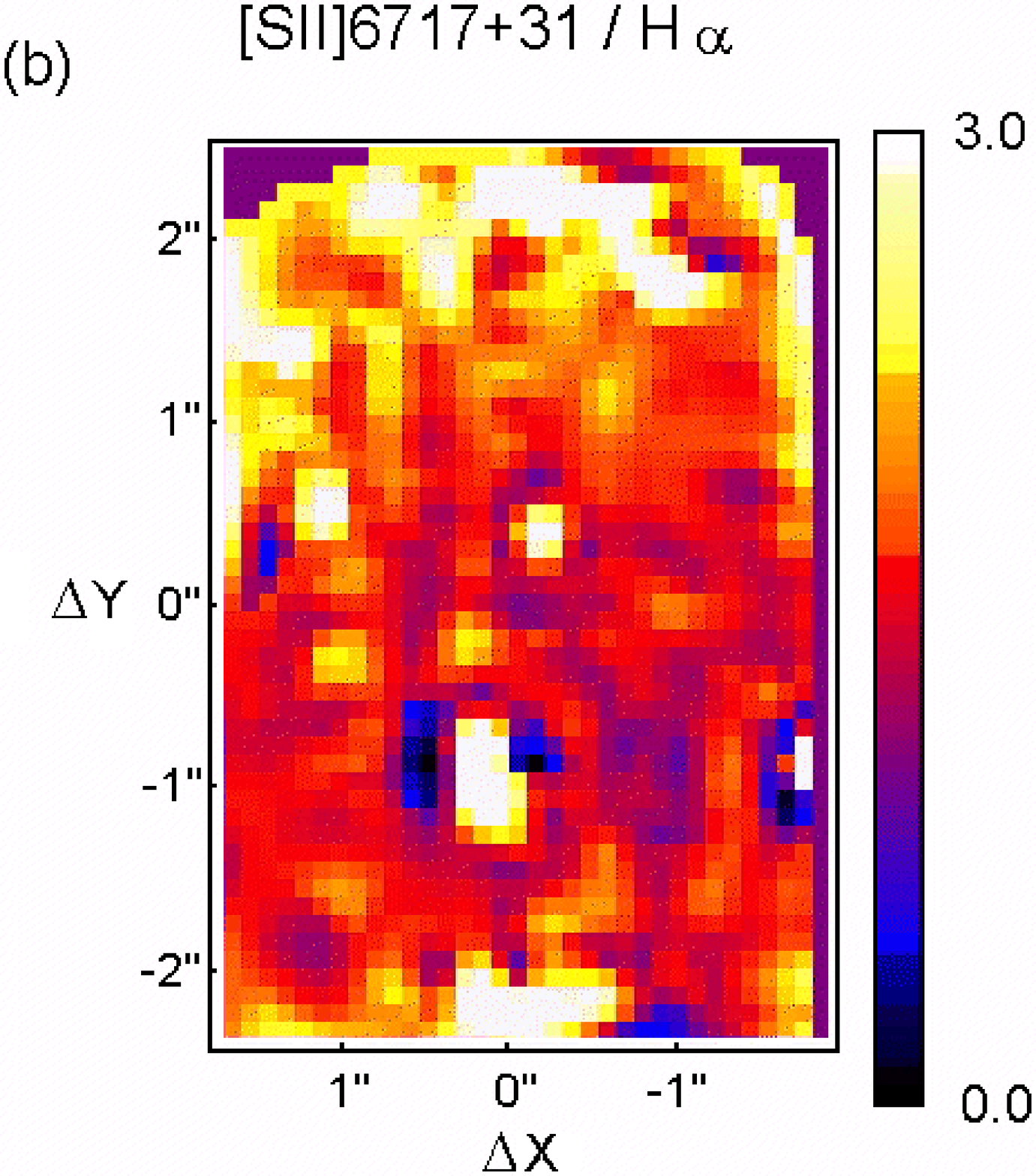}
  \caption{{\it Left:} HST WFPC2 Deep R Contour + GMOS field.\protect\\
    {\it Rigth:} [S {\sc ii}]/H$\alpha$ GMOS map.}
  \label{fig:ab1}
\end{figure}

{\bf Shell S3:}
Fig. 1a shows the very extended morphology of the main
super giant shell. This plot was performed using  a scale of
fluxes starting from very low values.
This very deep figure shows  almost the complete emission of S3, which
shows highly symmetric and circular shape, with center in the QSO.

Here, we remark some interesting properties found using the GMOS-3D
spectra, the emission line and kinematics maps:
(i) all the GMOS field --including S3-- depicts very extended emission line
and also blue continuum (at scale of $\sim$ 15-20 kpc around the QSO);
(ii) the knots of the shell S3 show multiple emission lines with Liner
properties, which are associated with shock processes. Furthermore,
the [S {\sc ii}]/H$\alpha$ map (Fig. 1b) shows the typical arc structures
associated with the external shocks, of the shells S3 and S2;
(iii) the kinematics of these emission lines is in agreement with 
extreme explosive/OF processes.

Lipari et al. (2007) already proposed that the GMOS results
obtained for shell S3 show propeties typical of: an expanding
 shell, and also of a companion/satellite galaxy.
This fact is in good agreement with the prediction of
theretical explosive models for formation of galaxies/QSOs (propossed
by Ikeuchi; Ostriker et al.).

{\bf Shells S1 and S2:}
The GMOS spectra, the physical condition diagrams and the kinematics maps
show also very interesting properties of these 2 internals
shells: multiple emission lines with Liner properties (associated with shocks);
plus the kinematics is consistent with this explosive scenario.



\section{Conclusion about IRAS 04505-2958 and the Explosive Model.}

{\bf 3.1. IRAS 04505-2958 and the concept of Galaxy Remnant/End.}

A very interesting point about IRAS 04505-2958 is the fact that though
shell S3 is clearly observed, at the same redshift
the host galaxy of this QSO remains undetected, in spite of the very careful
image analysis.
Multiple explosive events expelling a high fraction of the host galaxy could
be a possible explanation.
Several explosive events can eject a high fraction of the ISM; this process
would play a main role in the evolution of the star formation, finally
defining the mass of the remnant of the original galaxy.
We call a galaxy remnant this end product of the multiple explosive
processes. We believed that in IRAS 04505-2958 we are observing for
the first time a candidate of a galaxy in the end phase (via explosions),
and/or a remnant.


{\bf 3.2. Cosmic Rays Associated with AGNs/QSOs:
the new data from the P. Auger Observatory.}
Recently an important result was
obtained at P. Auger (Abraham et al. 2007). They found that
the cosmic rays (CR) with very high energy  are associated with
AGNs.
There are mainly two Models of AGNs/QSOs, which could explain these
observations:
(i) Obscured and Collimated AGN/Black-Hole; (ii) Evolutionary, Explosive,
and Composite Model.

The production of relativistic electrons is in
young SN remnants and it is believed that remnants simultaneously produce
relativistic ions/CRs (see Ellison et al. 2007).
On the other hand,
in the evolutive model for AGNs, HyN explosions  are a main
component; thus we suggest that giant remnants of HyN explosions (that
we call RHyN) could be a natural candidate for the origin --in AGNs--
of very energetic CRs.
In addition,
it is important to note that also the large duration and very energetic
gamma ray bursts are associated mainly with HyN explosions.


{\bf 3.3. Dark Matter in IR Mergers and BAL + IR + FeII QSOs.}
There are two interesting points about the relation of these
BAL QSOs and dark matter:
First, the sequence of these QSOs --in the
evolutionary IR colour-colour diagram-- start in IR Mergers
with strong OF. Our detailed 3D spectroscopic kinematical
study of IR Mergers + OF (like NGC 3256, NGC 2623, etc) show
mainly sinusoidal radial velocity curve. We already discussed the
possibility that the dark matter is absent in these IR Mergers
(Lipari et al. 2004).

Second, it is important to remark that in the last decades the
neutrinos were considered as a probable candidate for dark
matter, but it is not clear their origin. Recently,
the discovery of the most luminous SNe: 2006GY, 2006TF, 2005AP,
which are powered by the death of extremely massive star
strongly suggest that the neutrinos generated by HyN --and
primordial HyN-- could be considered as a candidate for the origin
of at least part of the dark matter.

\end{document}